# Observation of stable Néel skyrmions in Co/Pd multilayers with Lorentz transmission electron microscopy

Shawn D. Pollard[1*], Joseph A. Garlow[2,3*], Jiawei Yu[1], Zhen Wang[4], Yimei Zhu[3], Hyunsoo Yang[1]

1. Department of Electrical and Computer Engineering, National University of Singapore, 117576, Singapore
2. Material Science and Engineering Department, Stony Brook University, Stony Brook, NY 11794, USA
3. Condensed Matter Physics and Material Science Department, Brookhaven National Laboratory, Upton, NY 11973, USA
4. Department of Physics and Astronomy, Louisiana State University, Baton Rouge, Louisiana 70803, USA

*These authors contributed equally to this work.

Correspondence and requests for materials should be addressed to Y.Z. (email: zhu@bnl.gov) or H.Y. (email: eleyang@nus.edu.sg).

Néel skyrmions are of high interest due to their potential applications in a variety of spintronic devices, currently accessible in ultrathin heavy metal/ferromagnetic bilayers and multilayers with a strong Dzyaloshinskii-Moriya interaction. Here we report on the direct imaging of chiral spin structures including skyrmions in an exchange coupled Co/Pd multilayer at room temperature with Lorentz transmission electron microscopy, a high resolution technique previously suggested to exhibit no Néel skyrmion contrast. Phase retrieval methods allow us to map the internal spin structure of the skyrmion core, identifying a 25 nm central region of uniform magnetization followed by a larger region characterized by rotation from in- to out-of-plane. The formation and resolution of the internal spin structure of room temperature skyrmions without a stabilizing out-of-plane field in thick magnetic multilayers opens up a new set of tools and materials to study the physics and device applications associated with chiral ordering and skyrmions.

**Introduction**

Magnetic skyrmions are topologically non-trivial chiral spin textures present in systems with a strong Dzyaloshinskii-Moriya interaction (DMI) and a spatial extent of only tens to a few hundred nanometers. The discovery of skyrmions in bulk non-centrosymmetric magnets and thin films has led to the proposal of a variety of skyrmionic devices, in which the skyrmion is used as the basic information carrier in next generation logic and memory devices[1–9]. Thin film heavy metal/ferromagnetic (HM/FM) bi- and multilayers have emerged as an ideal candidate for the development of such devices due to their tunability through a variation of the constituent components and relative layer thicknesses. Further, the spin Hall physics associated with the heavy metal underlayer allows for straightforward manipulation of the local magnetic state[2,10,11]. In these systems, strong interfacial spin-orbit coupling results in a large interfacial DMI (iDMI), characterized by the DMI constant, $D$. The iDMI acts as an effective symmetry breaking in-plane field, $H_{DMI}$, which can stabilize Néel walls of a fixed chirality[12]. For films that satisfy the condition $\frac{|H_{DMI}|}{H_k} > \frac{2}{\pi}$, where $H_k$ is the anisotropy field, a stable Néel skyrmion can form[13,14]. Néel skyrmions are characterized by the magnetization rotating from one out-of-plane direction to the other radially from the skyrmion center, where the sense of rotation is set by the skyrmion chirality. The spin structure of a Néel skyrmion is schematically illustrated in Fig. 1a, b.

Néel skyrmions have been experimentally observed at room temperature in a variety of the aforementioned bilayers. These skyrmions have been limited to sizes well over 100 nm, making them ill-suited for practical applications and, at these length scales, their claim as skyrmions is still controversial[10,15,16]. Furthermore, the total magnetic thickness in these systems is below 1.5 nm, posing challenges for thermal stability. Recently, Moreau-Luchaire *et al.* demonstrated sub-100 nm magnetic skyrmions in dipolar coupled Pt/Co/Ir multilayers, in which the assymetric Pt and Ir layers result in an enhanced $H_{DMI}$ strong enough to generate skyrmions in films as thick as 6.6 nm[11]. However, to date, no skyrmions have been observed in thick ferromagnetic multilayers in which the disperate magnetic layers are exchange coupled. Furthermore, previous Néel skyrmion observations required either the presence of an out-of-plane bias field provided by either an external field or through dipole coupling to an adjacent ferromagnetic layer,[10,11,15–17] or careful tuning of a geometric confining potential to stabilize the skyrmion structure[18]. Cobalt/palladium (Co/Pd) is a prototypical exchange coupled multilayer which shows strong spin-orbit coupling effects in the bulk as evidenced by the large spin-orbit torque (SOT) efficiencies in nanowire geometries[19,20]. However, in HM/FM/HM systems, the contributions to the DMI from top and bottom interfaces is predicted to cancel out for identical HM layers[11,21,22].

In this work, we have imaged the formation of room temperature Néel skyrmions in a symmetric 6.7 nm thick Co/Pd multilayer with Lorentz transmission electron microscopy (L-TEM) for the first time. Importantly, the size of the observed skyrmions is significantly smaller than previously reported systems in which non-multilayer films are used[15,16,23], and of nearly identical extent as those measured in Ir/Co/Pt multlayers[11]. Following nucleation with an external magnetic field, the observed skyrmions persist even after removal of the external field and without geometric confinement. Furthermore, the film studied in this work is unique in that skyrmions are observed in an exchange coupled system with a large magnetic volume despite utilizing symmetric interfaces, i.e. Pd at both top and bottom interfaces for each layer as well as Pt layers at both outer surfaces. This work opens up a new set tools in which skyrmions can be designed, tuned, and studied for use in possible spintronic devices.

**Results**

**Imaging Néel skyrmions with L-TEM**

In comparison to other techniques such as Kerr microscopy and X-ray methods, which are limited by spatial resolution, L-TEM affords the advantage of a spatial resolution below 5 nm[24–26]. Furthermore, the scattering cross-section for electrons is significantly larger than that of X-rays[27], allowing for improvements in signal contrast in the ultrathin films which are typical of skyrmion systems. The contrast formed in L-TEM is traditionally explained as the result of the deflection of the electron beam due to an in-plane magnetic field, which leads to either increased or decreased contrast at a region of varying magnetization. For Néel-type domain boundaries in materials with perpendicular magnetic anisotropy (PMA), the Lorentz deflection follows the length of the wall, and suggests that no magnetic contrast is observed when the film is normal to the beam propagation direction. Using the full electron-wave treatment of the electron beam within the small defocus limit[28,29], the contrast in L-TEM resulting from the underlying magnetic structure is expressed by the curl of the magnetization along the beam propagation axis, given by

$I(\mathbf{r}, \Delta) = 1 - \frac{\Delta e \mu_0 \lambda t}{h}(\nabla \times \mathbf{M}(\mathbf{r})) \cdot \hat{\mathbf{z}}$, where $I(\mathbf{r}, \Delta)$ is the normalized intensity, $\Delta$ the defocus, $e$ the electron charge, $\mu_0$ the vacuum permeability, $\lambda$ is the electron wavelength, $h$ Planck's constant, $t$ the film thickness, and $\mathbf{M}(\mathbf{r})$ the film magnetization. The valid defocus range is determined by the region which the contrast transfer function is linear with defocus, and is dependent on a variety of microscope specific parameters such as beam divergence angle[29]. For a Néel skyrmion, the curl of the magnetization lies completely in the plane of the sample (Fig. 1a-d). Hence, with zero sample tilt, a Néel skyrmion produces no contrast[10] and therefore, while used extensively in the imaging of Bloch skyrmions[4,5,30], L-TEM has not been utilized to observe Néel skyrmions. However, by tilting the sample with respect to the beam direction, a projection of the curl on the beam axis proportional to the degree of tilt can be obtained, enabling skyrmion contrast. This contrast corresponds to the projection of the magnetization that is directed out of the sample plane orthogonal to the beam direction.

Fig. 1e-j shows a simulated tilt series of Néel skyrmion contrast in L-TEM[31,32]. As an input, a micromagnetically simulated Néel skyrmion of radius nominally 100 nm was used. It is apparent from the magnetization curl that the observed signal asymmetry is independent of the direction by which the magnetization rotates from the bulk state to the skyrmion center, therefore the chirality itself cannot be determined unless the sign of the DMI is known *a priori*. In the tilted geometry, L-TEM is sensitive to the Néel skyrmion polarity, defined as +1 when the magnetization at the skyrmion center points up with respect to the film, and −1 when pointing down. This is in contrast to the signal obtained for a Bloch skyrmion, in which case L-TEM is sensitive to the helicity (i.e. the sense of rotation of the skyrmion magnetization from the $+z$ to $-z$ direction), and not the polarity[33–35]. The contrast asymmetry of the Néel skyrmion reverses depending on the sign of the tilt, and disappears for zero tilt. At a given tilt angle, the largest contrast asymmetry occurs where the in-plane component of the magnetization curl is perpendicular to the tilt axis. Therefore, the distance between the point of minimum and maximum intensity represents the width of the projected skyrmion in the plane normal to the beam propagation direction.

**Experimental observation of skyrmions in Co/Pd multilayers**

To image nanoscale domain features with *in-situ* L-TEM, including skyrmion structures, MgO (2 nm)/Pt (4 nm)/Co (0.7 nm)/[Pd (0.5 nm)/Co (0.7 nm)]$_5$/Pt (2 nm) thin films, where the 5 denotes the layer repetition number as shown in Fig. 2a, were grown on both a 0.5 mm thick Si substrate for magnetization and structural characterization measurements as well as on 100 nm thick $Si_3N_4$ electron transparent membranes for TEM imaging and *in-situ* magnetization experiments. High resolution scanning TEM (STEM) was performed on a cross section of the deposited film and showed distinct Co/Pd layers (Fig. 2b). Electron energy loss spectroscopy (EELS) further confirmed the well separated nature of the respective Co and Pd layers (Fig. 2c). The films exhibit PMA at room temperature with an effective anisotropy, $K_{u,eff} = 0.24$ MJ m$^{-3}$ and a saturation magnetization, $M_s = 880$ kA m$^{-1}$, as determined by vibrating sample magnetometry (VSM, Fig. 2d, e). For the determination of $M_s$, the Pd layer thickness is included in total magnetic volume as hybridization of the *d*-shell electrons at the Co/Pd interfaces results in a non-negligible magnetic moment in the Pd layer[20]. Prior to L-TEM imaging, the magnetization was saturated along the beam propagation direction (defined as the −*z* direction, Fig. 2f).

L-TEM imaging was performed at a defocus of −1.2 mm, at room temperature, and zero field[36]. A tilt series taken at −15, 0, and 15 degrees is shown in Fig. 3a-c, respectively. The contrast, identical to that of Néel skyrmions shown in Fig. 1h-j, indicates the presence of skyrmions stable at room temperature. The intensity asymmetry shows a polarity of +1, as expected from the initialization procedure. A line profile across the structure is shown in Fig. 3d, with a distance between maximum and minimum intensities giving a width of 90 nm. The size allows us to conclude that these nanoscale domains are, in fact, magnetic skyrmions. At this length scale, dipolar energy alone has been suggested to be insufficient to stabilize domains of this size[11,37]. We show, in the following sections, that due to the large DMI in this system, these domains are stable and chiral, consistent with that of Néel skyrmions.

In order to directly relate the L-TEM images to the structure of the skyrmion as well as map its internal spin structure, we have performed a phase reconstruction utilizing the differential transport-of-intensity (D-TIE) method[38]. This method relates the change in phase of the electron beam due to the Aharonov-Bohm effect to the difference in contrast between two magnetic states.[25,39] We utilize an reference state of a saturated film along the $+z$ direction, and a final state of an isolated Néel skyrmion with a polarity of $-1$. Both images were acquired with a tilt of $-30$ degrees to the *x*-axis. The use of D-TIE in this case has the primary advantage of allowing for the removal of the electrostatic contribution to the TIE signal from the film and substrate. As a result, a Néel skyrmion would appear as a dipole like structure in D-TIE, with the projection of the magnetic induction appearing normal to the tilt axis. Furthermore, the total difference between the final and reference state also includes stray fields originating from the skyrmion core, as diagrammatically shown in Fig. 4a. Tilting the sample results in a projection of the magnetic structure and associated stray field into the plane normal to the beam direction. Hence, the phase reconstruction should show two flux closure loops in addition to a uniformly magnetized core region. The D-TIE reconstruction exhibits this structure, and confirms the presence and expected Néel skyrmion, as shown in Fig. 4b-e. From the phase reconstruction, we find the central core in which the spin structure is nearly uniform along the $+z$ direction extends nominally 25 nm, with the total extent of the skyrmion, inclusive of magnetization rotation, to be ~ 90 nm. It should be noted that, due to small thickness of these films, and hence weak magnetic signal, large defocus values were necessary. Large defocus values can artificially increase the apparent skyrmion size due to delocalization effects[29]. However, for the experimental conditions used here, this effect is expected to be minimal[40,41].

These nanoscale skyrmions were observed without the presence of an out-of-plane stabilizing field or geometric confining potentials, which were necessary in prior works. We find that stable zero field skyrmions in unpatterned thin films are possible within a narrow band of values of the DMI constant, $D$. We use micromagnetics to determine this range, as shown in Fig. 5, and obtain an approximate magnitude of $D$, assuming an exchange stiffness constant $A = 10.0 \pm 2.5$ pJ m$^{-1}$, consistent with the range of previous measurements in Co/Pd multilayers[42]. This results in an estimate of $|D| = 2.0 \pm 0.3$ mJ m$^{-2}$, similar to values obtained in other systems exhibiting a skyrmion structure. Outside of a narrow band of values of $|D|$, the skyrmion structure is no longer stable at small fields, as reported in ref [11], and shown in Fig. 5b-d.

In order to verify that only one chirality of skyrmion is stable for a given sign of $D$, we have further utilized micromagnetics to simulate two skyrmions with opposite chiralities but with an identical polarity by seeding a skyrmion with opposite signs of $D$. $M_s$, $K_{u,eff}$, and $A$ were fixed at 880 kA m$^{-1}$, 0.24 MJ m$^{-3}$, and 10 pJ m$^{-1}$, respectively. These states were then used as inputs to a single skyrmion state with a $D = 2.0 \pm 0.3$ mJ m$^{-2}$. While the seeded state corresponding to a $D > 0$ skyrmion does not change, the $D < 0$ skyrmion is unstable and transforms into a state identical to that of the initialized $D > 0$ skyrmion (Supplementary Figure 1 and Supplementary Note 1). This, coupled with the fact that skyrmions are stable only above a critical value of $D$ indicates that these structures must be chiral.

The large DMI found in our work is counter to previous reports, which suggest that the DMI strength in a symmetric multilayer stack is near zero as the contributions between the top and bottom Pd layers for each Co layer should cancel[11,21]. However, in Co/Pd multilayers, the large lattice mismatch between Co and Pd results in a lattice distortion that is 30% larger at the Co/Pd interface compared to the Pd/Co interface[43], breaking the symmetry of the two interfaces and providing a mechanism for a non-zero DMI. The inequivalence between the interfaces was previously suggested as the mechanism for the large SOT efficiencies found in Co/Pd multilayer nanowires[19].

To determine the sign of the *D*, which sets the chirality of the structures observed in our films, as well as rule out Pt as the origin for the strong positive DMI in the Co/Pd multilayer, we have performed polar Kerr microscopy to measure the asymmetric domain wall creep velocity[21,44] in a MgO (2 nm)/Pt (4 nm)/Co (0.32)/[Pd (0.34 nm)/Co (0.32 nm)]$_N$/MgO (2 nm)/SiO$_2$ (3 nm) multilayer, where *N* represents the multilayer repetition number which is varied from two to five, as shown in Fig. 6a-d. For the measurements, a semi-circular domain is nucleated around a pinning center in the as deposited film, while a field is applied 5° from the sample plane. Domain expansion is driven by the out-of-plane component of the field. The in-plane component breaks the rotational symmetry, a result of strong DMI in these films. The sign and strength of *D*, determined by this asymmetry (as detailed in Supplementary Note 2), is shown in Fig. 6e. For the $N = 2$ sample, we find $D < 0$, indicative of a large, negative DMI strength, as is expected for a Pt underlayer.[45,46] However, for $N > 2$, the asymmetry reverses, and a large $D > 0$ is found. This sign change can only be explained by a large, positive contribution to *D* from the Co/Pd multilayer structure, and allows us to further exclude Pt as the source of the strong positive DMI in this work[21,47].

**Magnetization reversal and the emergence of skyrmions**

To analyze the emergence of skyrmion ordering in the Co/Pd multilayer, and observe the role that iDMI plays in this process, the film magnetization was initialized in the +*z* direction, and then imaged while a magnetic field was applied nominally 30 degrees from the −*z* direction.

While a component of the magnetic field in the plane of the sample is present during the imaging process, it is not predicted to cause significant changes to the observed domain structure[48]. Figure 7 and Supplementary Movie 1 show the domain structure during *in-situ* application of a magnetic field opposite the initialization direction. After initial application of small fields, snake-like structures emerge with contrast equivalent to 360 degree Néel domains, with a width of nominally 100 nm (Fig. 7a) similar to the contrast expected for the domain pattern outlined in Fig. 5d. A simulated L-TEM image for Fig. 5d is shown in Supplementary Figure 2. Similar contrast has also been observed for homochiral Néel walls in thin film Pt/Co/AlO$_x$ samples in L-TEM, albeit with much larger domain spacings[12]. Minor increases in the field lead to the nucleation of a greater number of such domains (Fig. 7b, c), until the film becomes densely packed (Fig. 7d) making distinguishing individual domains difficult. This occurs over a narrow range of about 10 mT and corresponds to the initial drop in net magnetization seen in the VSM measurements in Fig. 2d.

Once formed, the domains are stable to significantly higher fields and do not merge with neighboring walls, providing evidence of their chiral nature, as argued in ref. [12]. For two walls of different chirality, there is an attractive interaction to each other, driving them to merge and create larger single domain regions, while the opposite is true for domain walls of the same chirality[49]. In order to annihilate Néel domain walls, the magnetization must rotate away from the easy axis defined by the iDMI. This is schematically shown in Fig. 7g. The size of the observed Néel domains are consistently below 200 nm, indicating there is not significant merging of neighboring domains and that the domains possess a single chirality, consistent with the effects of strong iDMI. The effect of these chiral domains on the macroscopic magnetization behavior is seen in the VSM measurement in Fig. 2d, where the magnitude of the slope of the reversal clearly decreases after the film becomes saturated with 360 degree Néel walls, indicating the presence of an intermediate state where the domain wall density reaches a maximum due to the large energy barrier necessary to form additional chiral Néel walls with spacing below a few hundred nanometers. Increasing of the field breaks the 360 degree Néel domains into smaller regions (Fig. 7e), eventually transforming into skyrmions before the saturation field (Fig. 7f). This results in a mixed FM/skyrmion state. The skyrmion state remains after removing the applied field.

To verify thermal stability, we have also initialized a state with a high skyrmion density, then waited three hours prior to subsequent imaging. No changes in the domain structure were observed, indicating the state is thermally stable at zero field. A note of caution here is warranted, as it appears that preferential domain propagation exists along the applied field direction in the early stages of the reversal. This is an artifact resulting from the lack of contrast associated with Néel walls along the tilt axis, and disappears when images are combined from both tilt-axis. This issue is discussed in Supplementary Note 3 as well as Supplementary Figure 3-4.

In summary, we have used L-TEM to track the formation of nanoscale Néel skyrmions in Co/Pd multilayers. The size and spin structure of the Néel skyrmion have been confirmed through phase reconstruction with D-TIE. Further, these Néel skyrmions are the first to be observed with zero bias field in an unpatterned film, an important step in the design of functional devices based on magnetic skyrmions. The presence of zero field stable skyrmions, coupled with measurements of asymmetric domain wall creep velocities in Co/Pd multilayers, has allowed us to estimate $D = 2.0 \pm 0.3$ mJ m$^{-2}$. Similar multilayers, which have already shown large SOT efficiencies in device geometries, merit further study in regards to skyrmion based logic and storage devices.

## Methods

**Thin film growth and imaging**

Thin film samples were grown via dc (metal layers) and rf (oxide layers) magnetron sputtering in an argon environment, with a base pressure of $7 \times 10^{-9}$ Torr. TEM measurements were performed at Brookhaven National Laboratory, using both a specially designed JEOL 2100F-LM and a JEOL ARM 200F microscope. Zero-field imaging was performed in the JEOL 2100F-LM, which has a residual out-of-plane field in the specimen area of less than 0.4 mT and a spherical aberration coefficient, $C_s$, of 109 mm. Reversal experiments were carried out in the JEOL ARM 200F, in which the out-of-plane magnetic field was tuned by varying the strength of the microscopes objective lens. The magnetic field at a given objective lens current was determined by mounting a Hall probe onto a TEM sample holder and measuring the Hall signal as the objective lens current was varied. Principal component analysis of the EELS line-scan has been smoothed with a Savitzky–Golay filter.

**Image simulations**

Simulated Lorentz images are taken from micromagnetically simulated skyrmion states in a 2 μm disc with a 4.2 nm thickness using object oriented micromagnetic framework (OOMMF) with a cell size of 2 nm × 2 nm × 4.2 nm cell, $D$ = 2.0 mJ m$^{-2}$, $K_{u,eff}$ = 0.24 MJ m$^{-3}$, $A$ = 10 pJ m$^{-1}$, and $M_s$ = 880 kA m$^{-1}$. The skyrmion was used as input for Lorentz image simulations using micromagnetic analysis to Lorentz TEM simulation (MALTS), with input parameters taken from the specs of the JEOL 2100F-LM used in this work and a defocus of −1 mm, at tilt angles varying from −30 to 30 degrees (Supplementary Movie 2). Micromagnetic simulations to obtain an estimate of $D$ were performed using the experimentally determined $M_s$ and $K_{u,eff}$. While $D$ and $A$ were varied over a range of values. A cell size equal to 2 nm × 2 nm × 6.7 nm and a damping of $\alpha = 0.5$ were used. More details, including how variations in anisotropy can affect the estimate of $D$, can be found in Supplementary Note 4 and 5, as well as Supplementary Figure 5.

**Acknowledgments**
The authors would like to thank Dr. Anthony Bollinger for technical assistance in magnetic field calibration of the JEOL ARM 200F, as well as Dr. Lijun Wu and Dr. Myung-Geun Han for useful discussions. This research was supported by the National Research Foundation (NRF), Prime Minister's Office, Singapore, under its Competitive Research Programme (CRP award no. NRFCRP12-2013-01) and the U.S. Department of Energy, Office of Basic Energy Sciences, under Contract No. DE-SC0012704. Research carried out in part at the Center for Functional Nanomaterials, Brookhaven National Laboratory, which is supported by the U.S. Department of Energy, Office of Basic Energy Sciences, under Contract No. DE-AC02-98CH10886.

**Author contributions**
S.D.P., H.Y., and Y.Z. conceived and designed the research. S.D.P. and Y.J. carried out thin film growth and TEM sample preparation. S.D.P. and J.A.G. acquired and analyzed the L-TEM data. Y.J. performed the characterization of multilayer films. Z.W. and J.A.G. performed STEM and EELS analysis. S.D.P. performed micromagnetic calculations and image simulations. S.D.P, J.A.G., and H.Y. prepared the figures and manuscript. All authors discussed the data and the results. H.Y. supervised the project.

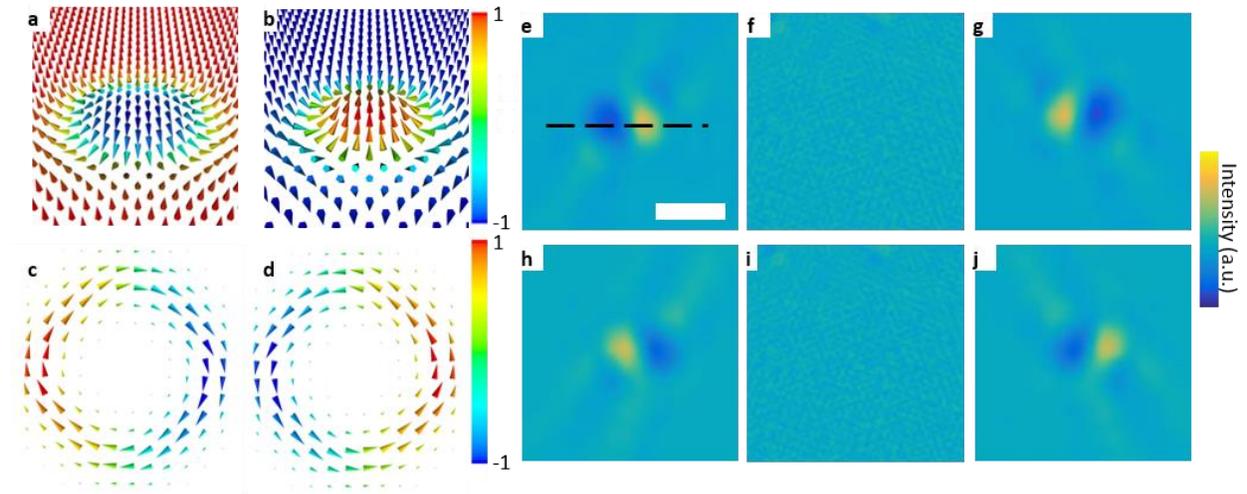

**Figure 1 | Simulated magnetic structure and L-TEM images of Néel skyrmions. a** and **b**, Spin structure of a skyrmion with positive DMI and a polarity of −1 and +1, respectively. The color corresponds to the normalized out-of-plane component of the magnetization. **c** and **d**, The corresponding curl of the magnetization from (**a**) and (**b**). The color represents the normalized *y*-component of the curl, which, when tilted about the *x*-axis, leads to a non-zero *z*-component, and results in the formation of Néel skyrmion contrast. **e**-**g**, Simulated Lorentz images at −30, 0, and 30 degrees tilt about the *x*-axis (indicated by the dashed line) for nominally 100 nm, skyrmion of −1 polarity taken at −1 mm defocus. **h**-**j**, Same, but for the spin structure a +1 polarity skyrmion. The scale bar is 200 nm.

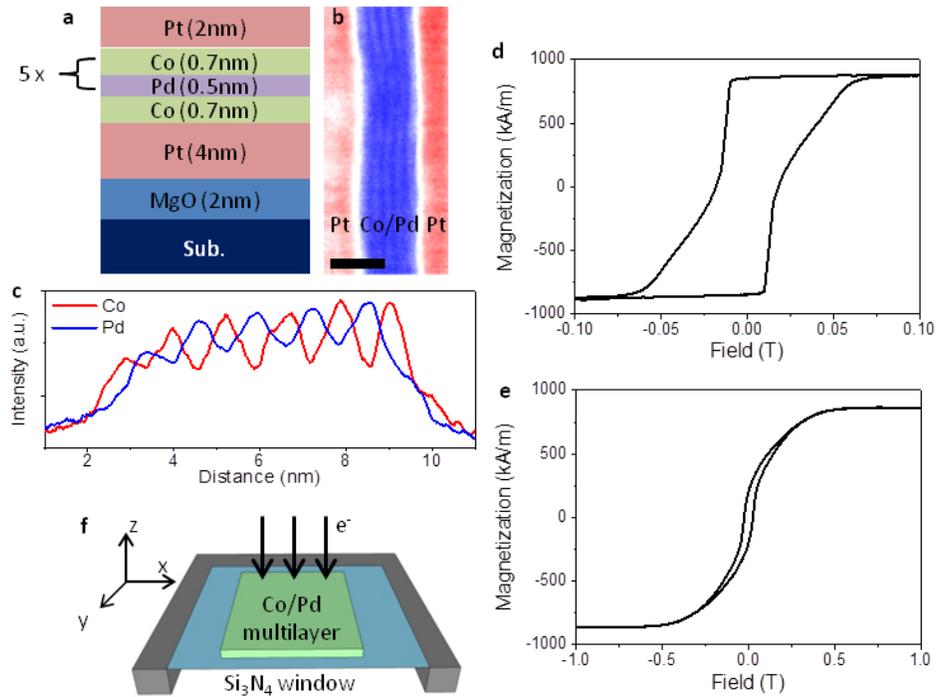

**Figure 2 | Sample geometry and bulk magnetic properties. a**, Schematic and **b**, STEM high angle annular dark field image of the thin film structure studied in this work. The scale bar is 5 nm. **c**, EELS line-scan showing the location of distinct Co and Pd layers. **d**, Out-of-plane and **e**, in-plane VSM hysteresis measurements of the film shown in (**a**), grown on a thermally oxidized Si wafer. The sloped reversal in (**d**) is indicative of domain mediated reversal. **f**, Experiment geometry for L-TEM imaging. The $Si_3N_4$ window is 0.5 μm × 0.5 μm and 100 nm thick.

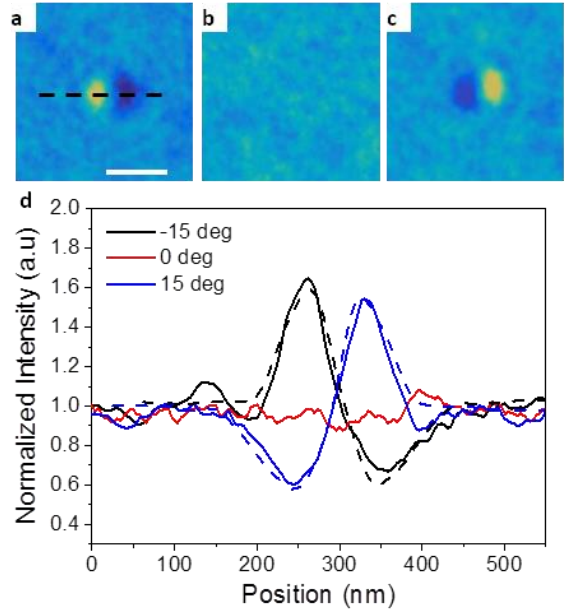

**Figure 3 | Experimental L-TEM images of a Néel skyrmion at varying tilt angles. a-c**, Tilt sequence of a magnetic skyrmion with a radius of 90 nm, taken at (**a**) −15, (**b**) 0, and (**c**) 15 degrees of tilt showing the disappearance of contrast at 0 tilt and reversal of contrast for opposite tilt angles. The tilt axis is indicated by the dashed line in (**a**). The scale bar is 200 nm. **d**, Line profiles of the three images, along direction of the dashed line in (**a**), showing the differences in contrast. The direction of asymmetry (bright-dark or dark-bright) is indicative of the skyrmion polarity. In this case, the imaged skyrmion polarity is +1. The skyrmion extent is determined by the distance between the maximum and minimum for the tilted samples. The dashed lines correspond to simulated intensity profiles of an 86 nm skyrmion, showing good qualitative agreement.

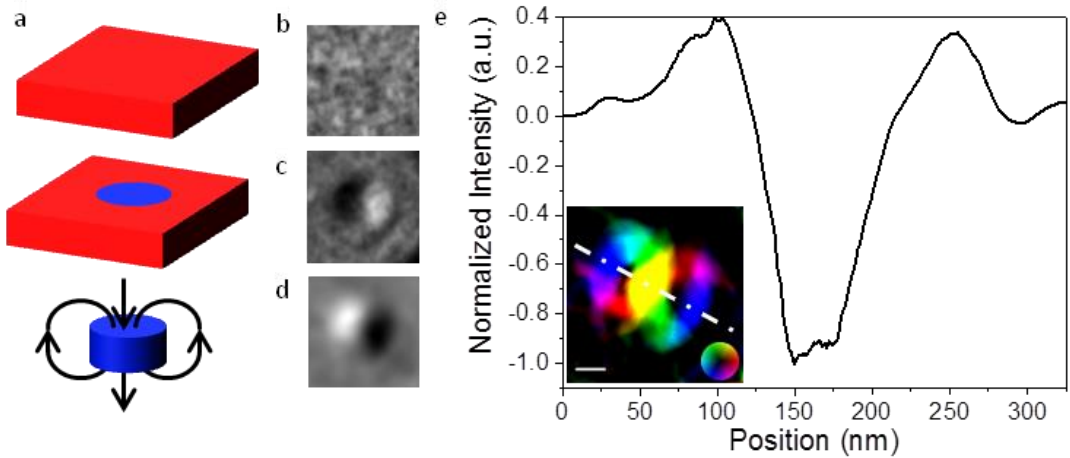

**Figure 4 | Magnetic induction map of a tilted skyrmion. a,** Schematic of the D-TIE reconstruction of a magnetic skyrmion. The red (blue) represents a region of magnetization along the $+z$ ($-z$) direction. A reference state (top) is used to remove the uniform magnetic background and electrostatic potential from a skyrmion state (middle), leaving the signal belonging solely to the skyrmion core and its associated dipole field. **b,** reference image corresponding to a uniform out-of-plane state. **c**, skyrmion image used for phase reconstruction. **d**, Reconstructed phase resulting from **b** and **c**. The images were taken at a negative defocus and −30 degree tilt. **e,** The component of the magnetic induction normal to the linescan indicated in the in-plane induction map is shown from an isolated skyrmion. The magnetic induction map showing induction normal to the beam propagation direction is represented in the inset. The linescan also indicates the tilt axis used during reconstruction. The color represents the direction of the magnetic induction, while the intensity represents the amplitude. The skyrmion core is defined as the region in which the intensity falls to zero, and is determined to be 90 nm, while the yellow central region indicates the region in which the magnetization uniformly points out of the sample plane and is nominally 25 nm in extent. The scale bar is 50 nm.
19

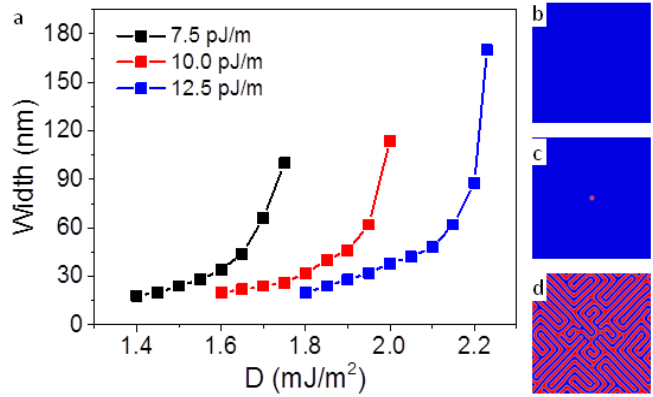

**Figure 5 | Simulated zero-field skyrmion width. a,** The simulated skyrmion width is shown at different values of *D* while varying *A*. Increasing *A* leads to a larger value of *D* necessary to stabilize the skyrmion. Further, larger values of *D* lead to a larger skyrmion size, until an upper limit is reached and the skyrmion deforms into a chiral stripe domain. Comparing to the experimental skyrmion width, we estimate a value of $|D| = 2.0 \pm 0.3$ mJ m$^{-2}$. **b-d,** Representative simulations of domain structures with *A* fixed at 10.0 pJ m$^{-1}$ and *D* (**b**) below, (**c**) within, and (**d**) above the critical range necessary to stabilize a skyrmion. Below the critical range, the exchange energy dominates and the skyrmion annihilates, while above this range, the skyrmion deforms into a labyrinthine stripe phase.

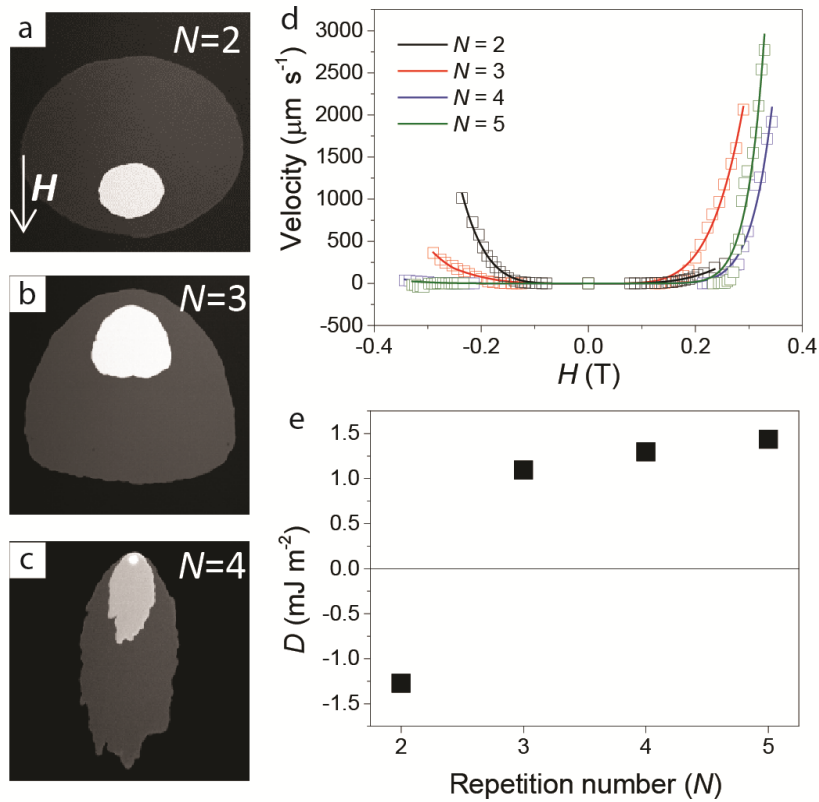

**Figure 6 | Determination of *D* via Kerr imaging of asymmetric domain wall creep. a-c,** Polar Kerr images showing the asymmetric domain wall expansion driven in a magnetic field tilted 5° from the indicated direction in (**a**). The asymmetry reverses sign from $N = 2$ to $N = 3$, where $N$ is the Co/Pd repetition number. **d,** Domain wall velocity vs. applied field. The solid line represents fits to a modified domain wall creep model which includes DMI contributions. **e,** *D*, extracted from the velocity curves in (**d**), in which a clear sign reversal between the $N = 2$ sample and $N = 3$ sample is observed, followed by a gradual increase in *D* for increasing the repetition number.

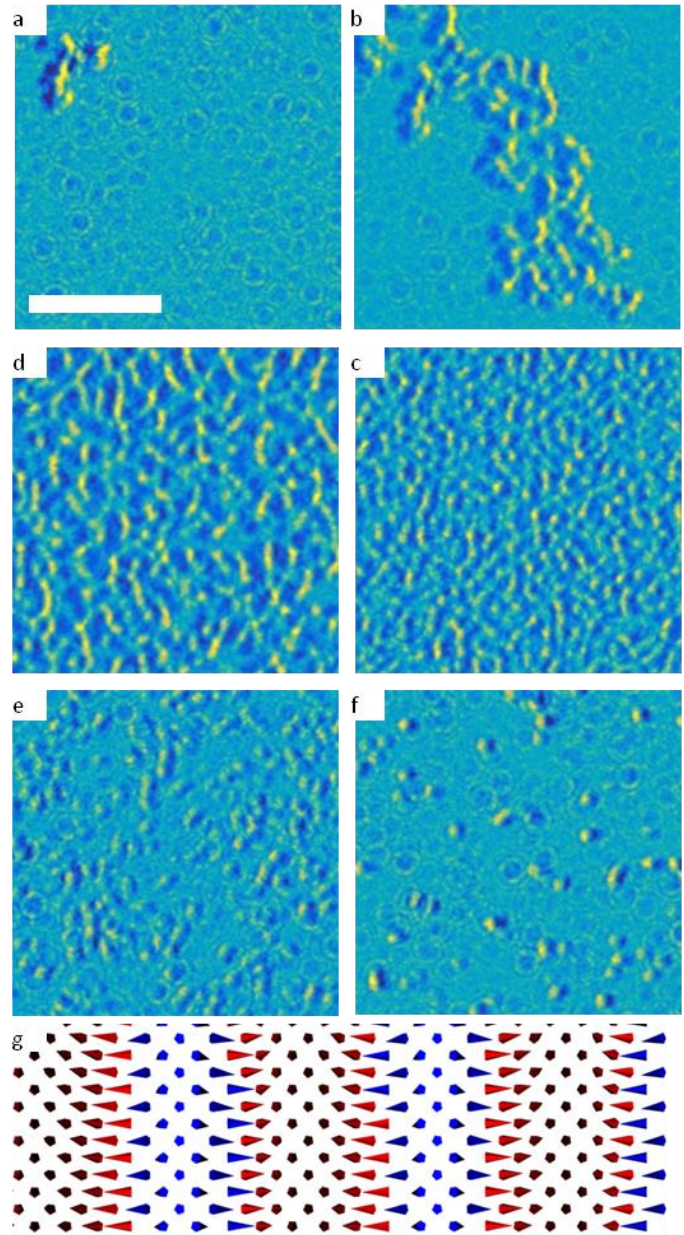

**Figure 7 | *In-situ* reversal behavior of a Co/Pd multilayer.** Partial reversal for the Co/Pd multilayer taken at (**a**) 11, (**b**) 16, (**c**) 23, (**d**) 50, (**e**) 72, and (**f**) 98 mT showing the nucleation and propagation of chiral Néel walls and the formation of skyrmions prior to saturation. Imaging was performed with a negative defocus and a −30 degree tilt. The tilt axis is along the horizontal direction. The scale bar is 2 μm. **g,** Schematic illustration of the magnetic structure of chiral 360 degree Néel domains. Red (blue) denotes the magnetization pointing along the +$z$ (−$z$) axis.